\documentclass[12pt]{iopart}
\usepackage[latin1]{inputenc}
\usepackage[english]{babel}
\usepackage{ragged2e}
\usepackage[final]{graphicx}
\usepackage{mathcomp}

\begin{document}

\title{Bilayer Graphene Quantum Dot Defined by Topgates}
\date{\today}
\author{Andr\'e M\"uller, Bernd Kaestner, Frank Hohls, Thomas Weimann, Klaus Pierz, and Hans W. Schumacher}

\address{Physikalisch-Technische Bundesanstalt, Bundesallee 100, 38116 Braunschweig, Germany.}
\ead{bernd.kaestner@ptb.de}

\begin{abstract}
We investigate the application of nanoscale topgates on exfoliated bilayer graphene to define quantum dot devices. At temperatures below 500$\,$mK the conductance underneath the grounded gates is suppressed, which we attribute to nearest neighbour hopping and strain-induced piezoelectric fields. The gate-layout can thus be used to define resistive regions by tuning into the corresponding temperature range. We use this method to define a quantum dot structure in bilayer graphene showing Coulomb blockade oscillations consistent with the gate layout.
\end{abstract}

\maketitle

Since the first synthesis of graphene~\cite{Novoselov2004} by exfoliation, the realization of quantum dots (QDs) in this material, e.g. as promising host for spin qubits~\cite{Trauzettel2007}, has been an interesting but challenging topic~\cite{Recher2010}. One of the outstanding challenges to fabricate reliable graphene QDs includes the difficulty in opening a sizeable and well-defined bandgap in graphene in order to define a QD confinement potential. It is possible to open a bandgap in three ways (see, e.g., \cite{Schwierz2010}): (1) by constricting the graphene in one dimension to form nanoribbons~\cite{Han2007, Sols2007, Stampfer2008c}, (2) by biasing bilayer graphene~\cite{maccann2006a, ohta2006, castro2007, oostinga2008, zhang2009} and (3) by applying strain to graphene~\cite{Ni2010, Pereira2009, Teague2009, Choi2010a, Verberck2012, Guinea2012}. 

Nanoribbon-QDs (1) produced by etching have been realized showing Coulomb blockade at sufficiently low temperature~\cite{Stampfer2008c} as well as single-electron pumping~\cite{Connolly2012}. A disadvantage of this method is that the required widths are lithography challenging and cause large carrier-mobility degradation. Another way to obtain nanoribbons utilizes unzipping carbon nanotubes which yield clean nanoribbon-QDs exhibiting Coulomb blockade, Kondo effect, clear excited states up to 20$\,$meV, and inelastic co-tunnelling~\cite{wang2011}. Following approach (2) QDs of arbitrary geometries and smooth tunable tunnel barriers may be achieved by local electrostatic gating of bilayers, as recently demonstrated by Allen~\emph{et al.}~\cite{yacoby2012a} in a dual-gate design. In their device a \emph{suspended} bilayer membrane has been placed between a global backgate and small topgates, not touching the membrane. With an appropriately shaped top-gate a Coulomb-island was defined and corresponding conductance oscillations through this QD have been observed. It would be highly desirable to obtain QD-devices on a \emph{substrate}, making the integration of more complex devices feasible. Recent success has been reported by Goossens \emph{et al.} where the graphene bilayer is sandwiched between hexagonal boron nitride bottom and top gate dielectrics~\cite{goossens2012}. In this work we follow a simpler approach and employ bilayer-graphene exfoliated on a GaAs-substrate. Due to leakage currents as a result of bonding we decide to use an undoped substrate and therefore pass on a backgate. A dielectric and metallic surface gates are deposited on top which are shaped to form a QD. At sufficiently low temperature we observe Coulomb blockade oscillations, consistent with the charging energy and the location of the transport channel expected from the gate layout. At even lower temperature the conduction is exponentially suppressed over the whole gate voltage range. We attribute this behaviour to resistive regions forming underneath the gate metal, leading to hopping transport at sufficiently low temperature. As the graphene is globally covered with a dielectric and not in direct contact with the metal gate we propose strain combined with strain induced piezoelectric fields inside the substrate as one of the possible origins for this behaviour.




\begin{figure}
	\centering
    \includegraphics{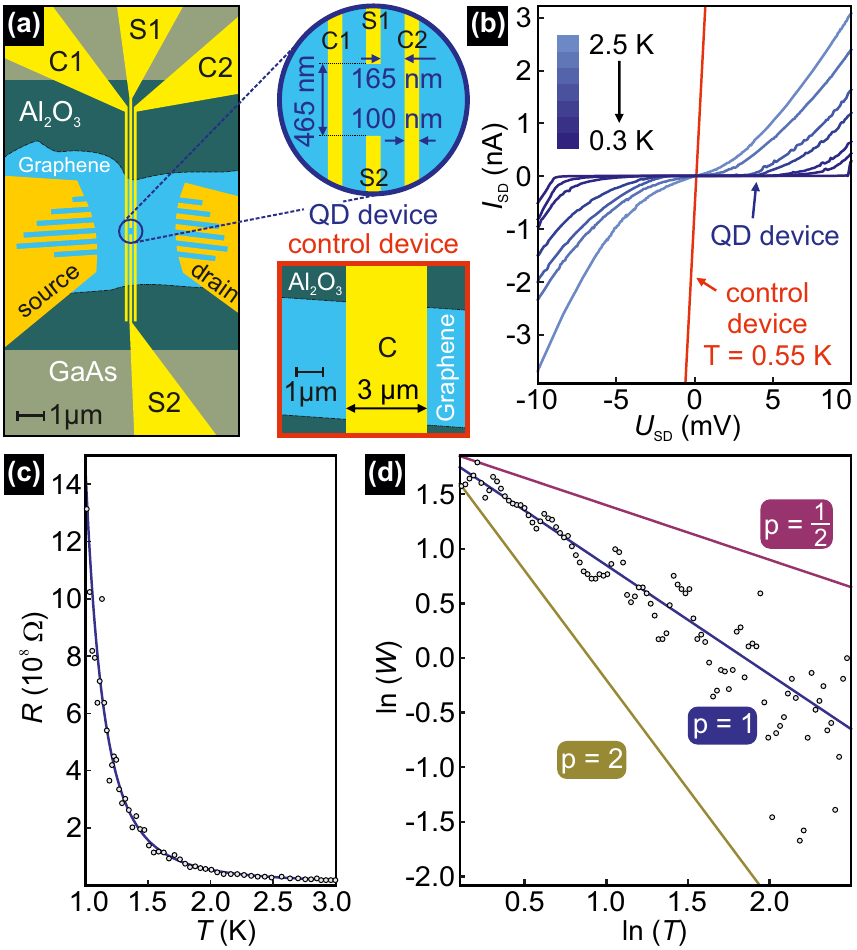}
    \caption{(a) Schematic of the QD and control device. (b) Source-drain current $I_\mathrm{SD}$ versus bias voltage $U_\mathrm{SD}$ at different temperatures, with all gates grounded. Characteristic of QD and control device are shown as blue and red traces, respectively. (c) Resistance $R$ of the QD device versus temperature with all gates grounded. The solid line corresponds to a fit to $R(T) = R_0 \exp \left( T_0 / T\right)^p$ with fit parameters $T_0, R_0$ and $p$ as described in the main text. (d) Reduced activation energy $W$ versus temperature $T$ on logarithmic scale, as explained in the main text. The solid lines indicate the slope corresponding to assumed $p$-values, as labeled.}
		\label{fig:Device}
\end{figure}

Fig.\ref{fig:Device}(a) shows a schematic of the device employed. The substrate of our device is undoped GaAs to increase the mobility at low temperatures due to the very low roughness, compared to the commonly used Si substrates with thermally oxidized SiO$_2$ on top~\cite{Woszczyna2011}. In order to allow optical separation between graphene of different layer numbers a 200$\,$nm thick GaAs/AlAs multilayer has been grown by molecular beam epitaxy onto the substrate which acts as a distributed Bragg reflector~\cite{Woszczyna2011, ahlers1009}. Following the exfoliation technique~\cite{Novoselov2004} a bilayer of graphene has been identified by optical microscopy and subsequent Raman spectroscopy. For the device a flake of about 20$\,\tcmu$m in length and 4$\,\tcmu$m in width has been chosen. It is contacted by pads of 10$\,$nm thermally evaporated titanium followed by 50$\,$nm gold. A dielectric is fabricated~\cite{Friedemann2012a} by first completely covering the whole device with an Al-layer of 2$\,$nm thickness by thermal evaporation. After venting the evaporation chamber, the thin Al layer is fully oxidized. It protects the graphene during the subsequent high-energy process step in which a 15$\,$nm thick gate dielectric of Al$_2$O$_3$ is sputtered. Finally, topgates are deposited by thermal evaporation of 10$\,$nm of titanium and 40$\,$nm of gold. In analogy to standard GaAs and recent suspended bilayer graphene structures~\cite{yacoby2012a} the gate geometry has been chosen to define a QD with dimensions as shown in Fig.~\ref{fig:Device}(a). The continuous gates are labelled C1 and C2, the split-gates S1 and S2, respectively. For comparison, a control device with a 3$\,\tcmu$m wide gate covering the bilayer has been fabricated. A voltage $U_\mathrm{SD}$ is applied between the contacts labelled ``source'' and ``drain''.


Fig.~\ref{fig:Device}(b) shows the source-drain current $I_\mathrm{SD}$ as function of bias voltage $U_\mathrm{SD}$ at different temperatures for the QD device (blue curves). For comparison, the characteristic for the control device at T = 550$\,$mK is shown by the red curve. All gate voltages are set to 0$\,$V. A non-linear characteristic with decreasing conductance is observed for the QD device as the temperature is lowered. The conduction is completely suppressed at the base temperature of $T \approx 300\,$mK below a threshold voltage for onset of conduction at $U_\mathrm{t} \approx \pm 9\,$mV. Note that this source-drain gap is not lifted for voltages of $\pm 1\,$V applied simultaneously to all gates. At higher voltages gate-leakage sets in. Hence, even without the application of an electric field the data show an apparent transport gap. This gap is not observed in the wide-gate control-device [red trace in Fig.~\ref{fig:Device}(b)]. Note, however, that at high voltages applied to the control device (see Fig. 2(b)) transport under the gate becomes suppressed and an electric field induced opening of a band gap and resulting nonlinear IV characteristics are found (not shown). Therefore, the conduction suppression is likely to be caused only when the surface gates are small and only within their proximity. Below we will discuss the assumption, that a combination of piezoelectric fields and strain due to a mismatch in thermal expansion coefficients of gate metal, insulating layer and GaAs substrate is the origin of the conduction suppression. The local strain can be much stronger in smaller structures of the same thickness (see supplementary data).

\begin{figure}
	\centering
    \includegraphics{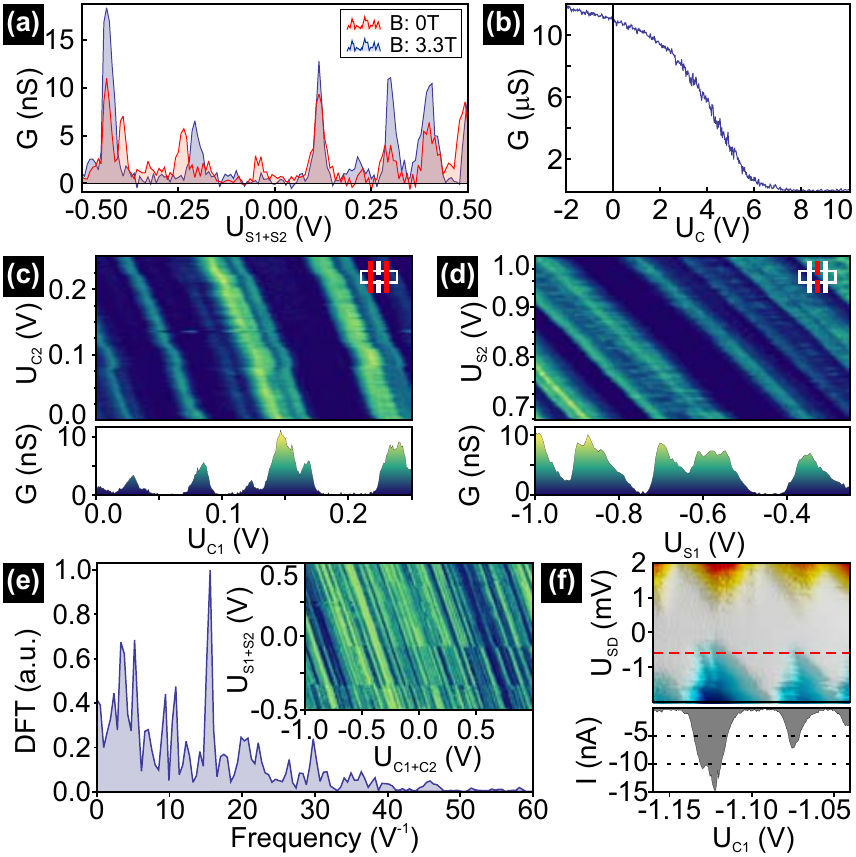}
    \caption{Coulomb blockade characteristic of QD device at temperature $T = 550\,$mK. (a) Conductance $G$ versus $U_\mathrm{S1+S2}$, for perpendicular magnetic fields of $B = 0$ (red) and $B=3.3\,$T (blue), respectively. The magnetic field was set to $B = 3.3\,$T in the following plots. (b) Conductance of the control-device versus gate voltage $U_C$. (c) and (d) show the conductance versus different gate voltages, as labeled. The line-cut at the bottom corresponds to the lowest voltage on the vertical axis. (e) displays the power spectrum obtained from a discrete Fourier transform of $G(U_\mathrm{C1+C2})$, averaged over $U_\mathrm{S1+S2} = -0.3 \ldots + 0.3\,$V. The inset shows the conductance $G$ as function of $U_\mathrm{C1+C2}$ and $U_\mathrm{S1+S2}$. The upper part of (f) displays the current $I_\mathrm{SD}$ as function of bias voltage $U_\mathrm{SD}$ and gate voltage $U_\mathrm{C1}$ with all other gates grounded. The bottom part shows the current $I_{SD}$ for the cut indicated by the dashed red line in the upper panel.}
	\label{fig:CBO}
\end{figure}

A conduction suppression only in the vicinity of the surface gates would leave a small island conducting in the center of the gate structure, which defines the QD. However, in order to allow probing the QD by transport measurement the temperature has to be raised to allow measurable transport through the structure. Fig.~\ref{fig:CBO} shows a corresponding measurement at $T = 550\,$mK. Here the conductance $G$ is shown as function of voltage applied equally to gates S1 and S2, $U_\mathrm{S1} = U_\mathrm{S2}$. A series of conduction peaks can be seen. We find that the application of a perpendicular magnetic field generally enhances the on-off ratio of these peaks (shown as blue trace), keeping the peak spacing approximately unchanged. For comparison, Fig.~\ref{fig:CBO}(b) shows the conduction-variation of the wide-gate control-device as function of gate voltage $U_\mathrm{C}$, also measured at $T = 550\,$mK. The conduction reduces smoothly towards the charge neutrality point, showing the typical characteristic of gated bilayer devices. No resonance or Coulomb blockade features have been observed. For voltages beyond 10$\,$V gate-leakage sets in which prevented the observation of carrier-type change. From this plot we conclude that the transport in the gated graphene layer is determined by holes.

Figs.~\ref{fig:CBO}(c) and (d) plot the conduction as the pairs of gate voltage ($U_\mathrm{C1}$, $U_\mathrm{C2}$) and ($U_\mathrm{S1}$, $U_\mathrm{S2}$) are varied, respectively. Apparently, all gates couple to the transport channel so that the total-conductance determining channel must be \emph{localized and close to the center of the gate structure}. This is a strong indication that resistive regions are forming underneath the gate metal, confining the transport to the QD region. It can also be seen that while gates S1 and S2 couple approximately equally to the transport channel, C1 and C2 show a different coupling. The ratio of the corresponding gate capacitance is $C_\mathrm{C1}/C_\mathrm{C2}\approx 4.2$ and $C_\mathrm{S1}/C_\mathrm{S2}\approx 0.84$. 

A discrete Fourier analysis of the conduction trace is shown in Fig.~\ref{fig:CBO}(e). It has been carried out for the conductance $G$ as function of simultaneously varied $U_\mathrm{C1}$ and $U_\mathrm{C2}$, over a range of $U_\mathrm{C1+C2} = -1 \ldots +1\,$V. The series of power spectra was calculated for fixed voltages $U_\mathrm{S1 + S2}$ simultaneously applied to gates S1 and S2, over a range of $U_\mathrm{S1 + S2} = -3 \ldots +3\,$V. The inset of Fig.~\ref{fig:CBO}(e) shows the corresponding $G(U_\mathrm{S1 + S2}, U_\mathrm{C1+C2})$-plot, and the main graph displays the average over all Fourier power spectra. A pronounced peak at a frequency of $17\,\mathrm{V}^{-1}$ can be seen, with a corresponding period of $\Delta U_\mathrm{C1+C2} \approx 59\,$mV.

Fig.~\ref{fig:CBO}(f) shows the variation of the current through the device as function of $U_\mathrm{C1}$ and voltage applied between source and drain, $U_\mathrm{SD}$. The current variation shows a typical Coulomb blockade oscillation (CBO) with a charging energy extracted from the Coulomb-diamond of $E_\mathrm{C} \approx 2\,$meV. The resulting total capacitance $C_\mathrm{T}$ of the corresponding Coulomb island is then $C_\mathrm{T} = e^2 /E_\mathrm{C} \approx 80\,$aF. From the self-capacitance model one would obtain a disc-shaped Coulomb island with a diameter of $d_\mathrm{island} \le C_\mathrm{T}/ 4 \epsilon_0 \epsilon_r \approx 220\,$nm, using a dielectric constant $\epsilon_r = 10.4$, which is the average between $\epsilon_\mathrm{GaAs} = 12.9$ of GaAs and $\epsilon_\mathrm{Al_2O_3} = 7.9 $ of Al$_2$O$_3$~\cite{Friedemann2012a}. The size estimated from this simple model is consistent in order of magnitude with the size of the gate defined island. Assuming that the main peak in the Fourier spectrum of Fig.~\ref{fig:CBO}(e) corresponds to Coulomb blockade oscillations of a single island the gate capacitance $C_\mathrm{C1+C2} = e / \Delta U_\mathrm{C1+C2} \approx 2.7\,$aF. Using the ratio $C_\mathrm{C1+C2}/C_\mathrm{S1+S2} = 2.25$ extracted from the data in the inset of Fig.~\ref{fig:CBO}(e) one obtaines a total gate capacitance of $C_\mathrm{G} \approx 3.9\,$aF.


In order to investigate the nature of the interface between Coulomb island and electron reservoir the temperature dependence of the conduction has been analyzed further in Fig.~\ref{fig:Device}(c) and (d). 
The voltage settings on the gates (all 0 V) correspond to a region where conductance is not suppressed by Coulomb blockade. Therefore the suppression of conduction for $|U_\mathrm{SD} | \le 9\,$mV at $T=300\,$mK [Fig.~\ref{fig:Device}(b)] originates from the interface, i.e. the region underneath the surface-gates. The temperature dependence can be described by the exponential law $R(T) = R_0 \exp \left( T_0 / T\right)^p$, where $T_0, R_0$ and $p$ are model specific constants. In order to determine the exponent $p$ we calculate the value from $\ln W = A - p \ln T$, where $W = -\partial \ln R(T)/\partial \ln T = p(T_0/T)^p$ is the reduced activation energy and $A$ is a constant~\cite{Khondaker1999}. The result can be seen in Fig.~\ref{fig:Device}(d) extracting $p = 1.1$, $R_0 = 2.9 \times 10^6\,$Ohm and $T_0 = 5.3\,$K. An exponent $p$ close to 1 suggests activated transport with an activation energy $E_a = 2 k_\mathrm{B} T_0 = 0.9\,$meV, with $k_\mathrm{B}$ Boltzmann's constant. 

Considering that the conduction at $T \approx 300\,$mK remains suppressed for voltages $U_\mathrm{C1+C2+S1+S2} = -1\,$V$ \ldots +1\,$V equally applied to all surface gates [see also Fig.~\ref{fig:CBO}(e) for CBO within $U_\mathrm{C1+C2} = -1\,$V$\ldots +1\,$V at $T = 550\,$mK] the 
corresponding transport gap must be significantly larger than the activation energy $E_a$. Therefore we interpret the simple activated behavior as a signature of nearest neighbor hopping (NNH) through localized states within the transport gap~\cite{Martin2009, Han2010}. Previously, NNH has been identified in bilayer graphene in which a band gap has been opened by a perpendicular electric field in a similar temperature range~\cite{Taychatanapat2010}. For our temperature study in Fig.~\ref{fig:Device}(c) and (d) no external perpendicular electric field has been applied that could justify the existence of a band gap~\cite{maccann2006a, ohta2006}. There might be a built-in electric field due to possible strong hole doping as indicated by the control-device in Fig.~\ref{fig:CBO}(b). However, this field is unlikely to lead to the QD definition according to the gate layout. It has previously been established that deformation -- described by a strain tensor $\varepsilon$ -- can induce a bandgap as well~\cite{Folger2008, Pereira2009a, Guinea2009}. The bandstructure of strained bilayer graphene has been theoretically investigated~\cite{Mariani2012, Verberck2012}, and a band gap has been predicted for perpendicular strain exceeding $\varepsilon_\mathrm{zz} = 0.25$. Also Choi~\emph{et al.}~\cite{Choi2010a} predict a gap when the two layers of bilayer graphene are differently strained. To support our arguments we perform simulations of the deformation and the forces built up inside the material described by a stress tensor $\sigma$. Indeed, at the substrate-graphene-dielectric interface of our sample both strain types exist.


\begin{figure}
	\centering
    \includegraphics{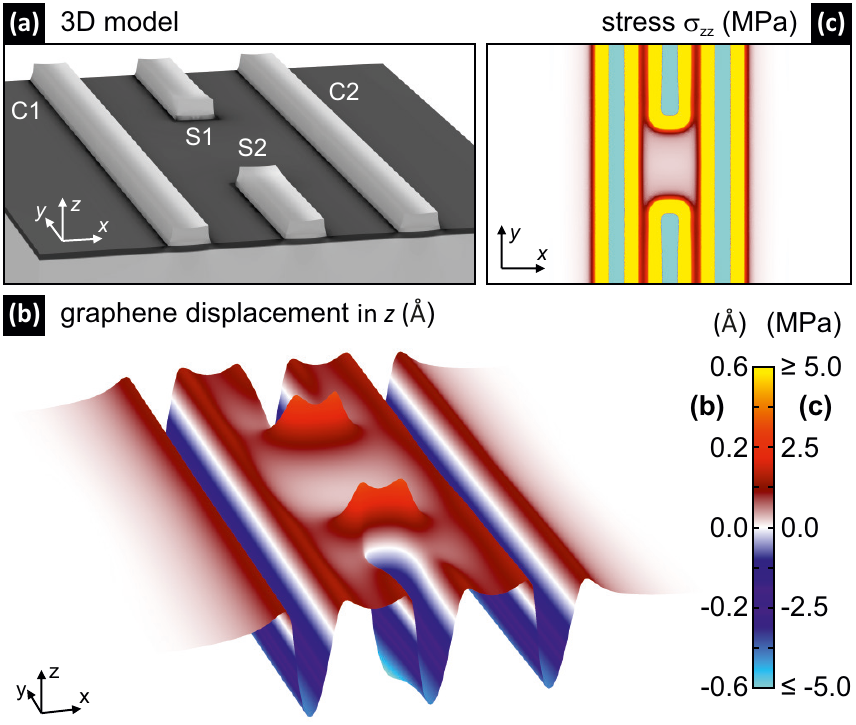}
    \caption{Simulation of sample-deformation at $T = 300\,$mK, using thermal expansion coefficients $\alpha_\mathrm{Au} = 14.2\cdot 10^{-6}\,$K$^{-1}$, $\alpha_\mathrm{Ti} = 8.6\cdot 10^{-6}\,$K$^{-1}$, $\alpha_\mathrm{\mathrm{Al}_2\mathrm{O}_3} = 6.5\cdot 10^{-6}\,$K$^{-1}$, and $\alpha_\mathrm{GaAs} = 5.7\cdot 10^{-6}\,$K$^{-1}$; assumed substrat temperature during deposition of Al$_2$O$_3$ was $T_\mathrm{dep} = 470\,$K; for Ti and Au $T_\mathrm{dep} = 290\,$K was assumed. (a) Overview of the 3D model and its deformation, scaled by a factor of 30. (b) plots the stress $\sigma_\mathrm{zz}$ and (c) the displacement in z of the graphene layer.}
		\label{fig:3D}
\end{figure}

The origin of deformation and internal forces in our sample is the fact that the insulator material and the metallic gates have different thermal expansion coefficients $\alpha$. The sample holder was kept close to room temperature during the evaporation of the surface gates. The temperature of the sample during the Al$_2$O$_3$ deposition was determined to be approximately $470\,$K. As the sample-temperature is changed the different materials will contract or expand at different rates and the deposited layers will excert a force on the substrate resulting in mutual deformation. The deformation in turn generates an internal elastic stress $\sigma$ that tends to restore the material to its original undeformed state.

The simulated deformation of the sample at $T = 300\,$mK is shown in Fig.~\ref{fig:3D}(a), with parameters provided in the caption. The simulation has been performed using a commercial finite element solver~\cite{comsol}. Plotting  the displacement of the GaAs-Al$_2$O$_3$ interface (i.e. the location of the graphene flake) in Fig.~\ref{fig:3D}(b) and (c) reveals that only a small rectangular island bordered by C1 and C2, as well as S1 and S2 remains almost free of displacement and therefore stress. In contrast, the graphene-portion covered by the gates is strongly displaced and therefore experiences strain due to internal forces of the surrounding material. In addition, the bending will result in a different strain for the two graphene layers.

\begin{figure}
	\centering
    \includegraphics{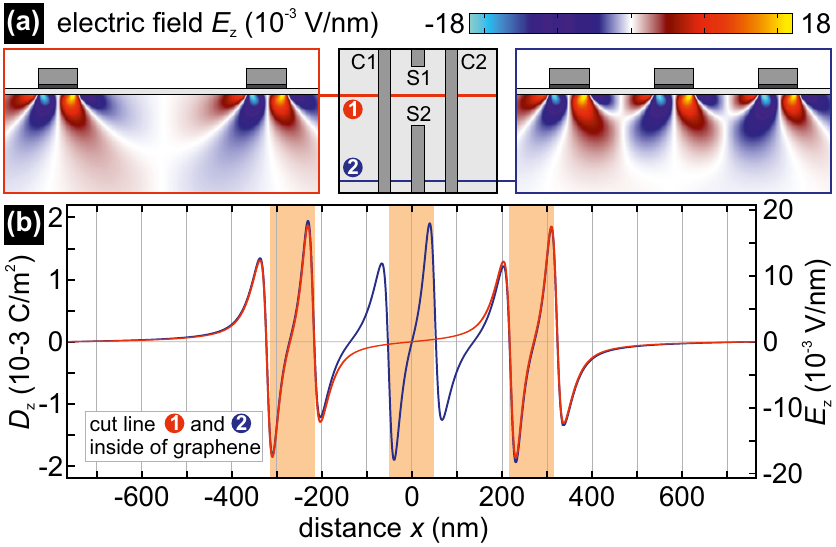}
    \caption{(a) illustrates surface plots of the simulated z-component of the electric field $E_\mathrm{z}$ in the x-z-plane at cut line 1 and 2, where two or three topgates are existent on top, respectively. (b) shows the corresponding electric displacement field $D_\mathrm{z}$ and the electric field $E_\mathrm{z}$ along the illustrated cut lines inside of the graphene. The orange boxes in the background depict the position of the top gates.}
		\label{fig:piezo}
\end{figure}

The stress was simulated for both a GaAs substrate and a Si substrate covered by SiO$_2$ and it turns out that the stress inside the graphene layer is in the same order of magnitude but slightly smaller in the case of Si/SiO$_2$. Concerning the latter graphene devices with comparable nanoscaled topgates have previously been investigated \cite{zhang2009,yacoby2012a,Taychatanapat2010}, but a transport gap has not been observed. Therefore the strain itself is not sufficant to clearly explain our results. In contrast to Si/SiO$_2$ in GaAs the strain will also lead to an electric displacement field $D$ due its piezoelectric properties. By taking them into account in the simulations the z-component of the electric displacement field $D_\mathrm{z}$ at the boundary between the insulator and our piezoelectric substrate can be derived and converted into an electric field $E_\mathrm{z}$ by the relation

\begin{equation}
	D_\mathrm{z}=\epsilon_\mathrm{0}\epsilon_\mathrm{r}E_\mathrm{z}.
\end{equation}

In Fig.~\ref{fig:piezo}(a) the calculated z-component of the electric field $E_\mathrm{z}$ is shown for two different sectional planes and in (b) for two section lines inside the graphene (see sample sketch). The maximum value of $18\,$mV/nm is reached beneath the gate edges and corresponds to a band gap opening of about $2\,$meV \cite{zhang2009}, which is in agreement with our experimental data.

In conclusion, we have investigated the functionality of nanoscale topgates on bilayer graphene exfoliated on a GaAs substrate. At low temperature the conductance through the gate covered graphene reduces exponentially even without the application of external voltages. We attributed this phenomenon to nearest neighbor hopping underneath the gate covered regions and propose a combination of strain and piezoelectric fields due to mechanical stress imposed by the gates above the graphene as a possible origin. For a certain temperature range allowing sufficient conduction Coulomb blockade dominated transport is observed, consistent with the nanoscale metallization layout. At present the low temperature range of our QD device is limited by the activated NNH transport through localized states underneath the gates. Here in the future a geometry optimized gate design as discussed in the supplement might allow to extend the operational temperature range of such strain defined devices.

This work has been supported by the SPP-1459 of the German Research Foundation (DFG). We acknowledge the help with insulator fabrication by R. Wendisch and the fruitful discussions with L. Schweitzer, B. Trauzettel, B. Verberck, P. Recher, C. Stampfer and T. Bj\"orkman.

\newpage
\renewcommand{\thepage}{A-\arabic{page}}
\renewcommand{\thefigure}{A-\arabic{figure}}
\section*{Appendix A: Supplementary Material}

\setcounter{page}{1}
\setcounter{figure}{0}

In this supplementary material we explore the prospects of strain engineering in a typical top-gate structure by exploiting different thermal coefficients and different device geometries. The application of Ti-Au top-gates as used in the main text for bilayer graphene is very common in conventional semiconductor structures, such as for defining nanostructures in GaAs electron gases. The possibility of stress and induced piezoelectric fields in the latter type has been considered previously, for instance in Ref.~\cite{Geisler2004}, where ideally it should be minimized. In the case of strain engineered graphene the opposite regime is desirable.

\begin{figure}
	\centering
    \includegraphics{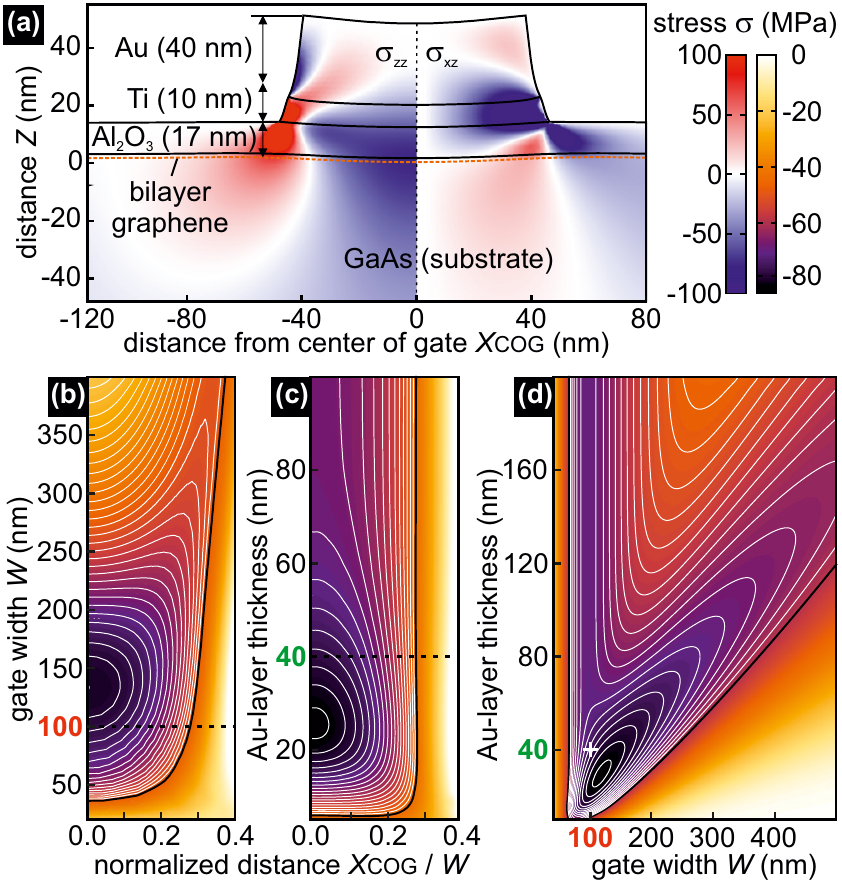}
    \caption{Simulation of the deformation and stress $\sigma$ at $T=300\,$mK of a single gate due to different thermal expansion coefficients. Geometrical paremters are provided in the cross-section in (a). The deformation has been scaled up by a factor of 30. The colors encode the stress distributions $\sigma_{zz}$ (left part of the gate) and $\sigma_{xz}$ (right part of the gate), as explained in the main text. The dashed orange line illustrates the location of graphene. (b) plots the stress felt by the graphene layer, $\sigma_{zz}$, as function of distance from the gate center and gate widths. The distance is normalized by the gate width $W$. (c) displays $\sigma_{zz}$ as the distance from the gate center is varied on the horizontal axis, for fixed $W=100\,$nm. On the vertical axis the gold-layer thickness was varied. The dashed lines indicate the actual experimental conditions. In (d) $\sigma_{zz}$ felt by the graphene layer under the center of the gate is plotted as function of gate width $W$ and gold-layer thickness. The white cross indicates the actual experimental condition. Between two contour lines $\sigma_{zz}$ varies by 2 MPa. From the black contour line on the subsequent lines are blanked out because of a too close spacing.}
		\label{fig:SingleGate}
\end{figure}

In Fig.~\ref{fig:SingleGate}(a) we evaluate the stress tensor $\sigma$ in the vicinity of the gates as used in the main text. It shows the cross-section of a single-gate with dimensions corresponding to our experimental setup and the deformation being graphically enhanced by a factor of 30. The corresponding stress profile of the two tensor components $\sigma_{zz}$ and $\sigma_{xz}$ are encoded in color. The location of the graphene bilayer is shown by the dashed line. For the normal component $\sigma_{zz}$ we find tensile stress ($\sigma_{zz} > 0$, red) near the gate corner, while towards the center there will be compressive stress ($\sigma_{zz}<0$, blue). In both cases strain induced band structure modifications and band gaps have been predicted~\cite{Verberck2012}. The right part of Fig.~\ref{fig:SingleGate}(a) shows the amount of shear stress $\sigma_{xz}$ obtained at the position of the bilayer. Note that band gap opening by shear strain has been predicted by Choi \emph{et. al.}~\cite{Choi2010a}. As discussed in the main text strain in piezoelectric substrates can also lead to significant local electric fields. This effect is not further detailed in this supplement.

Our simulation further shows that an increase in gate width $W$ relaxes the stress, as illustrated in Fig.~\ref{fig:SingleGate}(b). It shows the stress as function of position from the gate center, $X_\mathrm{COG}$, as its width is varied.  To allow comparison, the distance was normalized by the corresponding gate width $W$. Layer thicknesses according to the measured device in the main text have been used in the simulations [see Fig.~\ref{fig:SingleGate}(a)]. It can be seen that at $W = 350\,$nm the stress under the gate has relaxed to less than half of the maximum value at $W = 130\,$nm of $\sigma^\mathrm{max}_{zz} \approx - 80\,$MPa. The stress of $\sigma_{zz} \approx - 75\,$MPa in the middle of the gate for $W = 100\,$nm as used in our experiment lies close to the maximum (dashed horizontal line). These findings are consistent with the experimental results in the main text, where the wide-gate control device remained conducting at low temperatures. 

The above discussion suggests that sandwiching graphene between strained substrate-gate structures provides a complementary tool for defining graphene-nanodevices with tailored transport properties. An appropriate choice of material-sequence, evaporation technique, thickness, geometry etc. should allow tailoring the strain profiles for various applications~\cite{Guinea2012}. The wide opportunities of this technique can be derived from Fig.~\ref{fig:SingleGate}(d). For example, the stress can be maximized by choosing an optimum gold-layer thickness $t_\mathrm{Au}$ for a given gate width of $W = 100\,$nm, as seen in Fig.~\ref{fig:SingleGate}(c). A maximal value of $\sigma_{zz} = - 83\,$MPa is predicted for $t_\mathrm{Au} = 27.8\,$nm. The metal thickness in our experiment was $t_\mathrm{Au} = 40\,$nm, as shown by the horizontal dashed line. The relaxation for increasing $t$, however, saturates at $\sigma_{zz} = - 67\,$MPa for $t_\mathrm{Au} > 80\,$nm. This is expected as the stress relaxes for regions in large distance from the substrate.

Fig.~\ref{fig:SingleGate}(d) plots $\sigma_{zz}$ in the center of the gate as function of $t_\mathrm{Au}$ and $W$. It can be seen that a unique combination of $W=110\,$nm and $t_\mathrm{Au}=35\,$nm should globally maximize the stress. This example illustrates a possible procedure in finding optimal $t$ and $W$ parameters. The white cross in Fig.~\ref{fig:SingleGate}(d) indicates the stress value for the experimentally chosen parameters. The procedure may be extended to include arbitrary gate shapes, spacial thickness variations or material combinations in order to engineer a large variety of strain profiles.

\end{document}